\PassOptionsToPackage{square,numbers,sort&compress}{natbib}

\documentclass[final,fleqn,twocolumn,times]{elsarticle}

\usepackage{amsmath}
\usepackage{csquotes}
\usepackage{physics}
\usepackage{epstopdf}
\usepackage{mathtools}
\usepackage{cases}

\usepackage[capitalize]{cleveref}
\crefname{section}{Sec.}{Sec.}
\Crefname{section}{Section}{Sections}

\newcommand\restr[2]{{
  \left.\kern-\nulldelimiterspace 
  #1 
  \right|_{#2} 
  }}

\usepackage{jmr_arxiv}
\usepackage{framed,multirow}

\usepackage{amssymb}
\usepackage{latexsym}

\def\lunit{~\rm{\mu}m}
\def\tunit{~{\rm{ms}}}
\def\bunit{~{\rm{ms}}/\rm{\mu}m^2}
\def\dunit{~\rm{\mu}m^2/{\rm{ms}}}
\def\kunit{~\rm{\mu}m/{\rm{ms}}}
\def\gunit{~\rm{mT}/{\rm{m}}}

\usepackage{url}
\definecolor{newcolor}{rgb}{.8,.349,.1}
\usepackage[labelfont=bf,textfont=md]{caption}

\journal{Journal of Magnetic Resonance}

\begin{document}


\begin{frontmatter}

\title{The K{\"a}rger vs bi-exponential model: theoretical insights and experimental validations}

\author[1]{Nicolas \snm{Moutal}}
\ead{nicolas.moutal@polytechnique.edu}
\author[2]{Markus \snm{Nilsson}}
\author[2]{Daniel \snm{Topgaard}}
\author[1]{Denis \snm{Grebenkov}\corref{cor1}}
\ead{denis.grebenkov@polytechnique.edu}
\cortext[cor1]{Corresponding author:
Tel: +33 1 69 33 47 39
}

\address[1]{PMC, CNRS – Ecole Polytechnique, F-91128, Palaiseau, France}
\address[2]{Physical Chemistry, Lund University, P.O.B. 124, SE-22100 Lund, Sweden}


\begin{abstract}

We revise three common models accounting for water exchange in pulsed-gradient spin-echo measurements: a bi-exponential model with time-dependent water fractions, the K{\"a}rger model, and a modified K{\"a}rger model designed for restricted diffusion, e.g. inside cells. The three models are compared and applied to experimental data from yeast cell suspensions. The K{\"a}rger model and the modified K{\"a}rger model yield very close results and accurately fit the data. The bi-exponential model, although less rigorous, has a natural physical interpretation and suggests a new experimental modality to estimate the water exchange time.

\end{abstract}

\begin{keyword}
Water exchange; permeability; PGSE; yeast cells; K{\"a}rger model; bi-exponential model
\end{keyword}

\end{frontmatter}




\section{Introduction}

Diffusion Magnetic Resonance Imaging (dMRI) is a non-invasive technique allowing one to probe diffusion of nuclei in complex systems, in particular biological samples, with strong medical applications to brain and lung imaging \cite{Callaghan1991a,Price2009a,Grebenkov2007a,Tuch2003a,Frahm2004a,Bihan2012a}. In the case of free diffusion in a homogeneous medium, the classical Stejskal-Tanner formula for pulsed-gradient spin-echo (PGSE) sequences yields the signal $S=S_0\exp(-bD)$, where $S_0$ is the reference signal, $D$ is the diffusion coefficient of the spin-bearing particles in the medium and $b=\gamma^2 g^2 \delta^2 t_d$, where $\gamma$ is the nuclear gyromagnetic ratio, $g$ the magnetic field gradient, $\delta$ the gradient pulse duration and $t_d=\Delta-\delta/3$  
($\Delta$ being the delay between the two gradient pulses) \cite{Stejskal1965a}.
In biological systems, however, the restriction of diffusion by membranes or obstacles leads to a more complex dependence of the signal on the experimental variables $g$, $\delta$, $t_d$ \cite{Grebenkov2007a,Kiselev2017a}. A common strategy to fit non-exponential signal and to interpret the measurements consists in splitting the signal into two contributions: one coming from hindered diffusion in the extracellular space and the other from restricted diffusion in the intracellular space. The signal is thus fitted by a bi-exponential model, which yields the \textquote{fast} and \textquote{slow} apparent diffusion coefficients (ADC), respectively \cite{Niendorf1996a,Mulkern1999a,Clark2000a,Chin2002a,Sehy2002a,Ababneh2005a}.

The bi-exponential model is a convenient fit but its interpretations are not always correct \cite{Grebenkov2010b}. The distinction between free water and restricted water may be artificial \cite{Chin2002a,Schwarcz2004a,Kiselev2007a}. 
Moreover, the slow apparent diffusion coefficient often depends on the parameters of the gradient sequence, for example, the pulse duration $\delta$, that makes any comparison (and interpretation) between different experiments difficult or even impossible \cite{Mulkern1999a,Clark2000a}.
Finally, exchange of water between the intracellular and extracellular compartments may affect the result of the fit \cite{Stanisz2003a,Lee2003a}. It is worth stressing that the bi-exponential function is very flexible and allows to fit accurately a generic decaying signal. However, a good fit is not a definite \textquote{proof} of the underlying hypotheses of the model \cite{Grebenkov2010b,Novikov2010a}.

In order to account for exchange between the two pools of water, K{\"a}rger introduced in \cite{Kaerger1969a,Kaerger1971a} a model that was then developed to study diffusion NMR signals and  is in some sense an extension of the bi-exponential model \cite{Kaerger1985a,Kaerger1988a,Melchior2017a,Lauerer2018a}. The main idea consists in characterizing diffusion in the complex structure of the medium by macroscopic quantities, namely diffusion coefficients and exchange times. Fieremans \textit{et al} showed by Monte Carlo simulations that such a coarse-graining approach is valid in the regime of small cells and long exchange times.
\textcolor{black}{More explicitly, one has the conditions
\begin{equation}
\sqrt{Dt} \gg l_c \;, \qquad \text{and} \qquad \sqrt{D\tau} \gg l_c\;,
\label{eq:condition_Karger}
\end{equation}
namely,  the diffusion length $\sqrt{Dt}$ should be much larger than the correlation length of the medium $l_c$ and, at the same time, the exchange time $\tau$ should be much longer than the exploration time ${l_c}^2/D$ (``barrier-limited exchange'')  \cite{Fieremans2010a,Nilsson2013a}. This allows one to treat any complex medium as a ``homogeneous'' one where the exchange takes place at every point in space, which is the fundamental hypothesis of the K{\"a}rger model.}

The K{\"a}rger model originally relied on the narrow-pulse approximation (NPA) which is typically non valid for restricted diffusion inside compartments of a few microns if the encoding duration $\delta$ is greater than a few milliseconds. In \cite{Coatleven2014a,Li2014a} the K{\"a}rger model was rigorously extended to finite pulses, but the resulting ordinary differential equations need to be solved numerically. 
A \textquote{modified} K{\"a}rger model in which the slow ADC is set to zero was also proposed in order to account for restricted diffusion \cite{Price1998a}. Note that the derivation in \cite{Coatleven2014a,Li2014a} yields the same modified K{\"a}rger model with zero intracellular ADC, see Eqs. (20-29) from Ref. \cite{Li2014a}.

In this article, we critically revise the derivation of these three models (bi-exponential, K{\"a}rger model and modified K{\"a}rger model) and then apply them to analyze pulsed-gradient stimulated spin-echo experiments with yeast cells. The K{\"a}rger model and the modified K{\"a}rger model are shown to be very close to each other in the relevant range of parameters, whereas the bi-exponential model exhibits some deviations at low gradients. All three models fit the data well and give access to the exchange time across the cell membranes.

\section{Models}

For the sake of clarity and being motivated by experiments with yeast cells, we consider a medium that contains spherical cells of radius $R$. The diffusion coefficient of extracellular water is denoted by $D_{e}$ whereas the diffusion coefficient of intracellular water is denoted by $D_i$. We stress that these are the \textquote{true} diffusion coefficients and not ADCs extracted from spin-echo signals. In the following we implicitly assume that the spin-echo signals are normalized by the reference signal at zero gradient, i.e. $S(g=0)=1$.

Under the Gaussian phase approximation (GPA) and in the absence of exchange, Neuman derived the decay of the intracellular signal $S_i$ \cite{Neuman1974a}:
\begin{align}
&S_i \simeq  \rho \exp(-D_s b)\;,\label{eq:motional_narrowing}\\
&D_s=\frac{4R^2}{\xi t_d}\sum_{n=1}^{\infty} \frac{1-\frac{1}{{\alpha_n}^{2}{\xi}}F_n(\xi,\Delta/\delta)}{{\alpha_n}^4({\alpha_n}^2-2)}\;,\label{eq:D_s_complet}\\
&F_n(\xi,\Delta/\delta)=1-e^{-{\alpha_n}^2\xi}+2e^{-{\alpha_n}^2\xi{\Delta}/{\delta}}\sinh^2({\alpha_n}^2\xi/2)\nonumber\;,
\end{align}
where $\rho$ is the intracellular water volume fraction, $\xi=D_i\delta/R^2$ and $\alpha_n$ are the zeroes of the derivative of the spherical Bessel function $j_1$: $\alpha_1\approx 2.08, \alpha_2\approx 5.94,\ldots$.
The coefficient $D_s$ is thus the apparent \textquote{slow} diffusion coefficient probed by NMR. However, it is important to note that $D_s$ cannot be interpreted as a measure of mean-squared displacement since it depends \textit{a priori} on $\delta$.
When ${\alpha_1}^2\xi \gtrsim 1$ one can rewrite \cref{eq:D_s_complet} with a very good approximation as:
\begin{equation}
D_s\approx\frac{16R^2}{175\xi t_d}\left(1-\frac{F_1(\xi,\Delta/\delta)}{{\alpha_1}^2 \xi}\right)\;.
\label{eq:D_s_mod}
\end{equation}
In the limit $\xi\to \infty$ one recovers the well-known motional narrowing formula \cite{Grebenkov2007a}:
\begin{equation}
D_s\underset{\xi\gg 1}{\approx}\frac{16R^4}{175D_i \delta t_d}\;.
\label{eq:D_s}
\end{equation} 
\textcolor{black}{Unlike Eqs. \eqref{eq:motional_narrowing} and \eqref{eq:D_s_complet} which require $qR\ll 2\pi$, with $q=\gamma g \delta$, in order to satisfy the GPA, one can use Eq. \eqref{eq:motional_narrowing} with Eq. \eqref{eq:D_s} under the much weaker condition $qR \ll \xi$.
In contrast, when this condition is not satisfied (i.e. $qR \gg \xi \gg 1$), the GPA fails, and the signal exhibits ``abnormal'' dependence on the $b$-value, such as $-log(S) \sim b^{1/3}$ in the localization regime \cite{Robertson1966a,Swiet1994a,Hurlimann1995a,Grebenkov2014b,Kiselev2018a}.}

However, in typical experiments, $D_i \sim 1 \dunit$, $\delta \sim 1-10 \tunit$ and $R\sim 1-5\lunit$ which makes the condition $\xi\gg 1$ difficult to achieve. Therefore one generally has to carefully check the validity of the GPA, especially for the small values of $\xi$ (i.e., $\delta$).

In the absence of exchange across cell membranes, the complete signal can then be written as:
\begin{equation}
S=(1-\rho)\exp(-D_f b)+\rho\exp(-D_s b)\;,
\label{eq:biexp_no_exchange}
\end{equation}
where $D_f$ is the apparent \textquote{fast} diffusion coefficient, which is smaller than the intrinsic $D_e$ because the extracellular diffusion is hindered by the cells. Moreover, $D_f$ may decrease slowly with $t_d$ as demonstrated in \cite{Swiet1996a,Sen2004a,Novikov2014a}. Note that this involves already an approximation because we reduce the complex problem of diffusion in the extracellular medium to an apparent diffusion coefficient, ignoring for example localization effects at the cell boundaries \cite{Stoller1991a,Swiet1994a,Hurlimann1995a,Grebenkov2014b}. Now we investigate the effect of exchange accounted via three models.

\subsection{Bi-exponential model with time-dependent water fractions}
\label{section:bi-exp_mod}
The most simple idea is to keep \cref{eq:biexp_no_exchange} but to consider time-dependent intracellular water fraction $\rho$. 

Let us think of the magnetic field encoded water inside one cell as a \textquote{marked} water, which has intracellular and extracellular concentrations $c_i$ and $c_e$, respectively (at the beginning, $c_e=0$). This relies on the assumption that $\delta$ is sufficiently short so that water molecules stay inside the cell during the encoding. We also assume that the leakage is slow so that the intracellular concentration $c_i$ is at all times homogeneous inside the cell (\textquote{pore equilibration} \cite{Callaghan1991b}). Finally, we neglect the re-entrance of water because of the dilution in the extracellular medium: $c_e \ll c_i$. Then during the diffusion time the net rate of leakage of this marked water is $\kappa A c_i$, where $\kappa$ is the permeability of the cell membrane and $A$ its area, which yields the differential equation
\begin{equation}
\frac{\mathrm{d}c_i}{\mathrm{d}t}=-\frac{\kappa A c_i}{V} \;,
\end{equation}
where $V$ is the volume of the cell, from which one gets the classic formula
\begin{equation}
c_i=c^0_i \exp(-t/\tau_{i\to e})\;, \quad \tau_{i\to e}=\frac{V}{A\kappa}\;.
\label{eq:c_decay}
\end{equation}
For a sphere of radius $R$ the expression of $\tau_{i\to e}$ can be simplified (and for an arbitrary shape of diameter $2R$ the result is always smaller): 
\begin{equation}
\tau_{i\to e}=\frac{R}{3\kappa}\;.
\label{eq:tau_kappa}
\end{equation}

We can now come back to our assumptions: (i) the encoding is sufficiently short to neglect the effect of permeability, i.e. $\delta \ll \tau_{i\to e}$; (ii) the intracellular medium is \textcolor{black}{homogeneous at all times}, i.e. $\tau_{i\to e}\gg R^2/D_i$ or $R\ll D_i/\kappa$. One recognizes on the right-hand side the \textquote{permeability length} which represents the typical distance traveled by a particle near a boundary before crossing it \cite{Sapoval1994a,Sapoval2002a,Grebenkov2006c,Grebenkov2006d}. As we are in the restricted diffusion regime $\delta \gtrsim R^2/D_i$ we only have to check the first hypothesis. \textcolor{black}{Note that this corresponds to the conditions of applicability of the K{\"a}rger model \eqref{eq:condition_Karger}, with $l_c=R$}.

The above reasoning can be extended to multiple cells. In this case one still considers the intracellular water as marked water, but with \textcolor{black}{individual} markings for each cell. Indeed, \cref{eq:motional_narrowing,eq:D_s_complet} are valid only for water that stays inside the same cell. 
The magnetization of water molecules that travels from cell to cell is destroyed in the same way as freely diffusing water (actually, this is only true if the cells are not regularly arranged on a lattice, that we implicitly assume here).
 
The above analysis implies that $\rho$ should decay with $t_d$ according to Eq. \eqref{eq:c_decay}:
\begin{equation}
\rho=\rho_0\exp(-t_d/\tau_{i\to e})\;.
\label{eq:rho_decay}
\end{equation}
This model is \textit{a priori} only applicable in the case when the extracellular diffusion rapidly destroys the magnetization, that is $q^2D_f\tau_{i\to e} \gg 1$. Indeed, otherwise one should also take into account the entry of extracellular water whose magnetization is not negligible.

\subsection{K{\"a}rger Model}

The classical model for treating exchange between two compartments with different diffusion coefficients is the K{\"a}rger model \cite{Kaerger1985a,Kaerger1988a}. Roughly speaking, this is an extension of the bi-exponential model with an additional parameter: an exchange time $\tau_K$ which is the time-scale of the leakage from one compartment to the other. More precisely,
\begin{equation}
\tau_{i\to e} = \rho \tau_K \qquad \text{and} \qquad \tau_{e\to i}=(1-\rho)\tau_K
\label{eq:tau_tauK}
\end{equation}
are respectively the mean times for crossing the membranes from the inside to the outside and from the outside to the inside. The K{\"a}rger model relies on the assumption that $\delta \ll \tau_K$, which allows one to neglect the effect of exchange during the encoding and decoding gradient pulses. This means that, as far as the exchange is concerned, one can use $t_d=\Delta-\delta/3$ instead of, say, $\Delta + \delta$, as the total time during which the exchange takes place. \textcolor{black}{As a matter of fact, in the case of long-exchange times, it was shown that using this form of $t_d$ as the total time improves the accuracy of the K{\"a}rger model to the first order in $\delta/\Delta$ \cite{Nguyen2015c,Ning2018a}. In addition, it makes the comparison with the bi-exponential model easier.}

Solving the system of differential equations on the intra- and extracellular signals
\begin{subnumcases}\displaystyle
\frac{\mathrm{d}S_i}{\mathrm{d}t}=-D_s q^2 S_i -  S_i/{\tau_{i\to e}} +S_e/{\tau_{e\to i}} \label{eq:equation_Karger_intra}\\ \nonumber \\ \displaystyle
\frac{\mathrm{d}S_e}{\mathrm{d}t}=-D_f q^2 S_e - S_e/{\tau_{e\to i}} +S_i/{\tau_{i\to e}}
\label{eq:equation_Karger_extra}
\end{subnumcases}
and the initial conditions
\begin{equation}
S_i(t=0)=\rho \;, \quad S_e(t=0)=1-\rho\;,
\label{eq:init_Karger}
\end{equation}
one gets the K{\"a}rger formula
\begin{equation}
S=P_1\exp(-D_1q^2 t_d)+P_2\exp(-D_2q^2 t_d)\;,
\label{eq:Karger}
\end{equation}
where $P_1,P_2,D_1,D_2$ are functions of $q$ given by
\begin{align*}
D_{1,2}&=\frac{1}{2}\left(X_e+X_i\mp\sqrt{(X_e-X_i)^2+\frac{4}{q^4\tau_{e\to i}\tau_{i\to e}}}\right)\;,\\
X_e&=D_f+\frac{1}{q^2\tau_{e\to i}} \;, \qquad X_i=D_s+\frac{1}{q^2\tau_{i\to e}}\;,\\
P_1&=\frac{D_2-\rho D_s-(1-\rho)D_f}{D_2-D_1}\;,\\
P_2&=\frac{\rho D_s + (1-\rho)D_f - D_1}{D_2-D_1}\;.
\end{align*}

Note that some authors \cite{Price1998a,Meier2003a} claim using $4$ initial conditions for the two first-order differential equations \eqref{eq:equation_Karger_intra} and \eqref{eq:equation_Karger_extra} even though only $2$ conditions are needed. In our notations, the two additional conditions are
\begin{equation}
\left.\frac{\mathrm{d}S_i}{\mathrm{d}t}\right|_{t=0}=-D_s q^2 \rho\;, \; \left.\frac{\mathrm{d}S_e}{\mathrm{d}t}\right|_{t=0}=-D_f q^2 (1-\rho)\;,
\label{eq:init_faux_Karger}
\end{equation}
which are actually equivalent to each other and compatible with \cref{eq:tau_tauK}. Although one can interpret these redundant initial conditions as another way to state Eq. \eqref{eq:tau_tauK}, it is more natural, from the mathematical point of view, to discard Eq. \eqref{eq:init_faux_Karger}, keeping the two initial conditions \eqref{eq:init_Karger} and two physical relations \eqref{eq:tau_tauK}.

\subsection{Modified K{\"a}rger model}

One obvious flaw of the K{\"a}rger model is that $D_s$, which was supposed to be a constant intrinsic diffusion coefficient, depends on the diffusion time (see Eq. \eqref{eq:D_s_mod}). Although it seems to be of no consequence in the final formula \labelcref{eq:Karger}, it is a serious issue when one looks at the original equation \eqref{eq:equation_Karger_intra}. Should one treat $D_s$ first as a constant and then add its time dependence in the final formula or on the contrary consider that it is time-dependent from the beginning? Another defect is that the K{\"a}rger model is not supposed to be valid in the restricted diffusion regime $\xi \gtrsim 1$. In this case, the equation for the intracellular signal should be modified.

Actually, if one goes back to \cref{eq:motional_narrowing}, one can see that the time-dependence of $D_s$ in Eq. \eqref{eq:D_s_mod} is simply another way to state that the intracellular signal does not depend on the diffusion time in the restricted diffusion regime. Thus one \textcolor{black}{can} modify the K{\"a}rger model in the following way, inspired by \cite{Price1998a}:
\begin{subnumcases}
\displaystyle
\frac{\mathrm{d}S_i}{\mathrm{d}t}=-S_i/{\tau_{i\to e}}+S_e/{\tau_{e\to i}}
\label{eq:equation_Karger_mod_intra} \\ \nonumber\\ \displaystyle
\frac{\mathrm{d}S_e}{\mathrm{d}t}=-D_f q^2 S_e - S_e/{\tau_{e\to i}}+S_i/{\tau_{i\to e}}
\label{eq:equation_Karger_mod_extra}
\end{subnumcases}
with the initial conditions 
\begin{equation}
S_i(t=0)=\alpha\rho\;\quad S_e(t=0)=1-\rho\;,
\label{eq:init_Karger_mod}
\end{equation}
where $\alpha=\exp(-D_s b)<1$ depends on $q$ and $\delta$ but not on $t_d$. Compared to the K{\"a}rger model, the intracellular ADC is set to zero and the initial condition for the intracellular signal is different. Here, $\alpha$ is the time-independent \textcolor{black}{decrease} of the intracellular signal computed by \textcolor{black}{Neuman formulas \eqref{eq:motional_narrowing} and \eqref{eq:D_s_complet}}. One can see that Eqs. \eqref{eq:equation_Karger_mod_intra} and \eqref{eq:equation_Karger_mod_extra} provide the correct solution in the absence of exchange ($\tau_{i\to e}, \tau_{e\to i} \to \infty$). 

The main physical motivation behind this model is that the intracellular magnetization reaches an equilibrium on a much shorter time-scale than the water exchange through the cell membranes ($R^2/D_i \ll \tau_K$). \textcolor{black}{Because it does not evolve after this very short transient regime (in the absence of exchange), the corresponding ADC is set to zero. The initial value $\alpha \rho$ that we set for the intracellular signal is precisely the value of the signal resulting from this transient regime.}

Solving Eqs. \eqref{eq:equation_Karger_mod_intra} and \eqref{eq:equation_Karger_mod_extra} yields
\begin{equation}
S=P'_1\exp(-D'_1q^2 t_d)+P'_2\exp(-D'_2q^2 t_d)\;,
\label{eq:Karger_mod}
\end{equation}
where $P'_1,P'_2,D'_1,D'_2$ are functions of $q$ given by
\begin{align*}
D'_{1,2}&=\frac{1}{2}\left(X'_e+X'_i\mp\sqrt{(X'_e-X'_i)^2+\frac{4}{q^4\tau_{e\to i}\tau_{i\to e}}}\right)\;,\\
X'_e&=D_f+\frac{1}{q^2\tau_{e\to i}} \;, \qquad X'_i=\frac{1}{q^2\tau_{i\to e}}\;,\\
P'_1&=\frac{D_2(1-\rho(1-\alpha))- (1-\rho)D_f}{D_2-D_1}\;,\\
P'_2&=\frac{(1-\rho)D_f - D_1(1-\rho(1-\alpha))}{D_2-D_1}\;.
\end{align*}
One can see that the formulas for $D'_1$ and $D'_2$ are the same as the ones from the K{\"a}rger model with $D_s$ set to zero. However, the formulas for $P'_1$ and $P'_2$ are different due to the change of initial conditions.

As in the previous section, we note that some authors \cite{Price1998a,Meier2003a} write $4$ initial conditions instead of $2$ for Eqs. \eqref{eq:equation_Karger_mod_intra} and \eqref{eq:equation_Karger_mod_extra}. Their two additional conditions read in our notations as
\begin{equation}
\left.\frac{\mathrm{d}S_i}{\mathrm{d}t}\right|_{t=0}=0\;, \quad \left.\frac{\mathrm{d}S_e}{\mathrm{d}t}\right|_{t=0}=-D_f q^2 (1-\rho)\;,
\label{eq:init_faux_Karger_mod}
\end{equation}
which are equivalent to each other but \emph{not} compatible with \cref{eq:tau_tauK}. Because these authors probably used the same initial conditions \eqref{eq:init_Karger_mod} as us for the derivations, their formulas are the same as ours. However, the additional conditions \eqref{eq:init_faux_Karger_mod} implicitly discard \cref{eq:tau_tauK}, which expresses the conservation of mass and is thus a fundamental relationship between exchange times and water fractions.
To avoid further confusion, \textcolor{black}{the incompatible conditions} \eqref{eq:init_faux_Karger_mod} should be discarded.
\subsection{Comparison of the models}

We have considered three different macroscopic models for the exchange between the intracellular and the extracellular water in the restricted diffusion regime. The bi-exponential model is the most simple and intuitive one, the K{\"a}rger model is the canonical one, whereas the modified K{\"a}rger model is the most rigorous of the three in this situation. It seems natural to ask whether these three models give similar or different results and under which conditions.

\textcolor{black}{First, it follows from the mathematical definition of the modified K{\"a}rger model that it coincides with the K{\"a}rger model in the limit $D_s/D_f \to 0$. However, from a physical point of view, the modified K{\"a}rger model makes sense only if $D_s$ is inversely proportional to $t_d$, which necessarily implies that $D_s \ll D_i$ (see Eq. \eqref{eq:D_s_mod}) and thus $D_s \ll D_f$.} As a consequence, when the K{\"a}rger model and the modified K{\"a}rger model are applicable, they generally yield results that are close to each other.

As for the bi-exponential model with decreasing fraction $\rho$, one can expand the K{\"a}rger model at high gradients and long exchange time ($D_f q^2\tau_K \gg 1$) to get:
\begin{subequations}
\begin{align}
D_1&\approx D_s+\frac{1}{q^2\tau_{i\to e}}\;,\\
D_2&\approx D_f\;,\\
P_1&\approx \rho\;,\\
P_2&\approx(1-\rho)\;,
\end{align}
\end{subequations}
that shows that the bi-exponential model is close to the K{\"a}rger model in this regime. \textcolor{black}{To see this, we treat separately the cases of short and long diffusion times.
At short times ($D_f q^2 t_d \lesssim 1$), the extracellular signal is not completely attenuated, but one has $t_d \lesssim (D_f q^2)^{-1} \ll \tau_K$ so that exchange can be neglected. In other words, one can use $D_1\approx D_s$ and $D_2\approx D_f$, which yields the standard bi-exponential model.
At long times ($D_f q^2 t_d \gg 1$), the extracellular signal is completely attenuated, and the total signal reduces to the intracellular part:
\begin{equation}
P_1\exp(-D_1 q^2 t_d) \approx \rho \exp(-D_s q^2 t_d) \exp(-\frac{t_d}{\tau_{i\to e}})\;,
\end{equation}
which again coincides with the bi-exponential model with variable water fractions.}
Discrepancies between the two models appear at low gradients ($D_f q^2 \tau_K \ll 1$), which is consistent with the remark at the end of \cref{section:bi-exp_mod}.

In the next section we apply these three models to experimental data on yeast cells to compare their quality and range of applicability.

\section{Material and Methods}

Baker's yeast (J{\"a}stbolaget, Sweden) was purchased at a local supermarket, diluted with tap water in approximate volume ratio 1:2 (yeast:water),  transferred to a 5 mm NMR tube,  stored in room temperature for four days, and finally centrifuged at 1500$g$ for 2 min to form a packed cell sediment of 2 cm height.
NMR experiments were performed on a Bruker Avance-II spectrometer operating at 500.13 MHz \textsuperscript{1}H resonance frequency. The magnet was fitted with a Bruker MIC-5 probe with 3 T/m maximum gradient at a current of 60 A. The \textsuperscript{1}H signal of water was recorded with a pulsed gradient stimulated echo sequence \cite{Tanner1970a} for an array of values of $q$, $\delta$, and $t_d$ \cite{Aaslund2008a,Aaslund2011a,Lasic2011a}.
More precisely, four values of $\delta$ were used: $3.0\tunit$, $5.6\tunit$, $10.6\tunit$, $20\tunit$, and six values for $t_d=\Delta-\delta/3$: $20.2\tunit$, $35.2\tunit$, $187.2\tunit$, $327.2\tunit$, $572.1\tunit$, $1000.2\tunit$, yielding $24$ different curves. To avoid spurious effects of differences in $T_2$ between the intra- and extracellular components \cite{Eriksson2017a}, the time duration for transverse relaxation was held constant at 44.8 ms for all measurements.

The variable $q=\gamma g\delta$ took $26$ logarithmically spaced values from $5.3\cdot 10^{-3}\lunit^{-1}$ to $1.4 \lunit^{-1}$ whatever $\delta$ and $t_d$.
\textcolor{black}{The parameter $b=q^2t_d$ reached maximum values} of about $40\bunit$ for $t_d=20.2\tunit$ and about $2000\bunit$ for $t_d=1000\tunit$. The signal was systematically renormalized by the value $S_0$ at $b=0$ obtained by fitting a single exponential function $S = S_0\text{exp}(-bD)$ to data points fulfilling $S/S_0>0.8$.
Before performing any fit, we determined the noise level of the data to be about $0.25\%$.  Because the signal never goes down below $3\cdot 10^{-2}$ we conclude that the signal-to-noise ratio is always bigger than $10$.

\section{Results}

The typical radius of the yeast cells is $2.5\lunit$. The smallest encoding duration $\delta$ is $3\tunit$ for which $(D_i\delta)^{1/2}\sim 2\lunit$, implying the restricted diffusion regime ($\xi \approx 1$), but not the motional narrowing regime ($\xi \to \infty$).
The advantage of being in this intermediate regime is that by fitting $D_s$ with \cref{eq:D_s_mod}, one \textcolor{black}{can estimate} the two physical quantities $R$ and $D_i$, that is not possible in the motional narrowing regime (cf. \cref{eq:D_s}) \cite{Aaslund2009b}.

\subsection{Bi-exponential model with decaying $\rho$}

We have applied the fit \labelcref{eq:biexp_no_exchange} to all the values of $\delta$ and $t_d$. The quality of the fit was assessed by the value of the residual error, which was very close to the estimated noise value, indicating a good fit. Moreover, the $95\%$ confidence intervals on the fit parameters were each time about: $\rho\pm 1\%$, $D_f\pm 2\%$, $D_s \pm 4\%$.

\begin{figure*}[tb]
\begin{center}
\includegraphics[width=50mm]{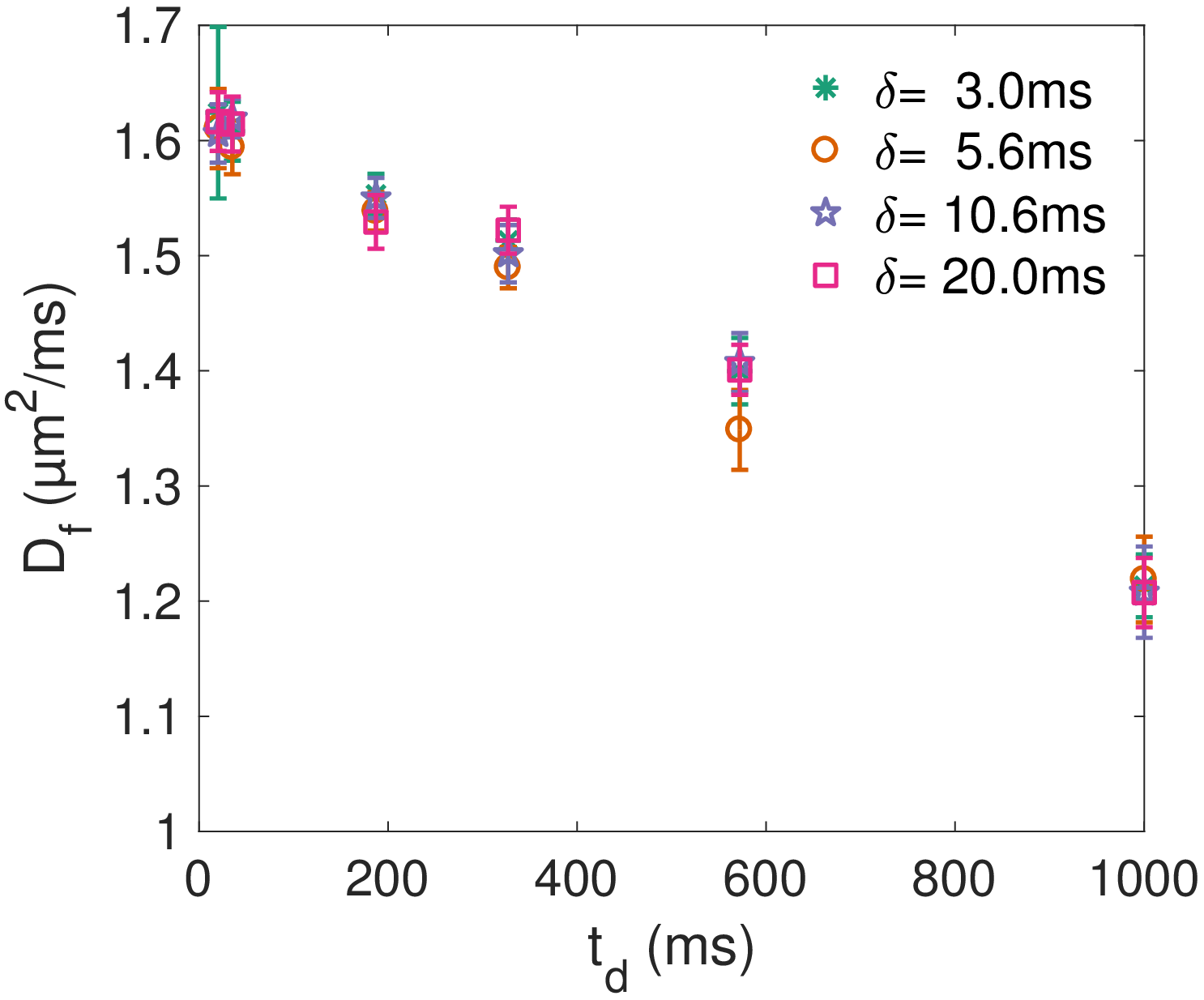}
\includegraphics[width=50mm]{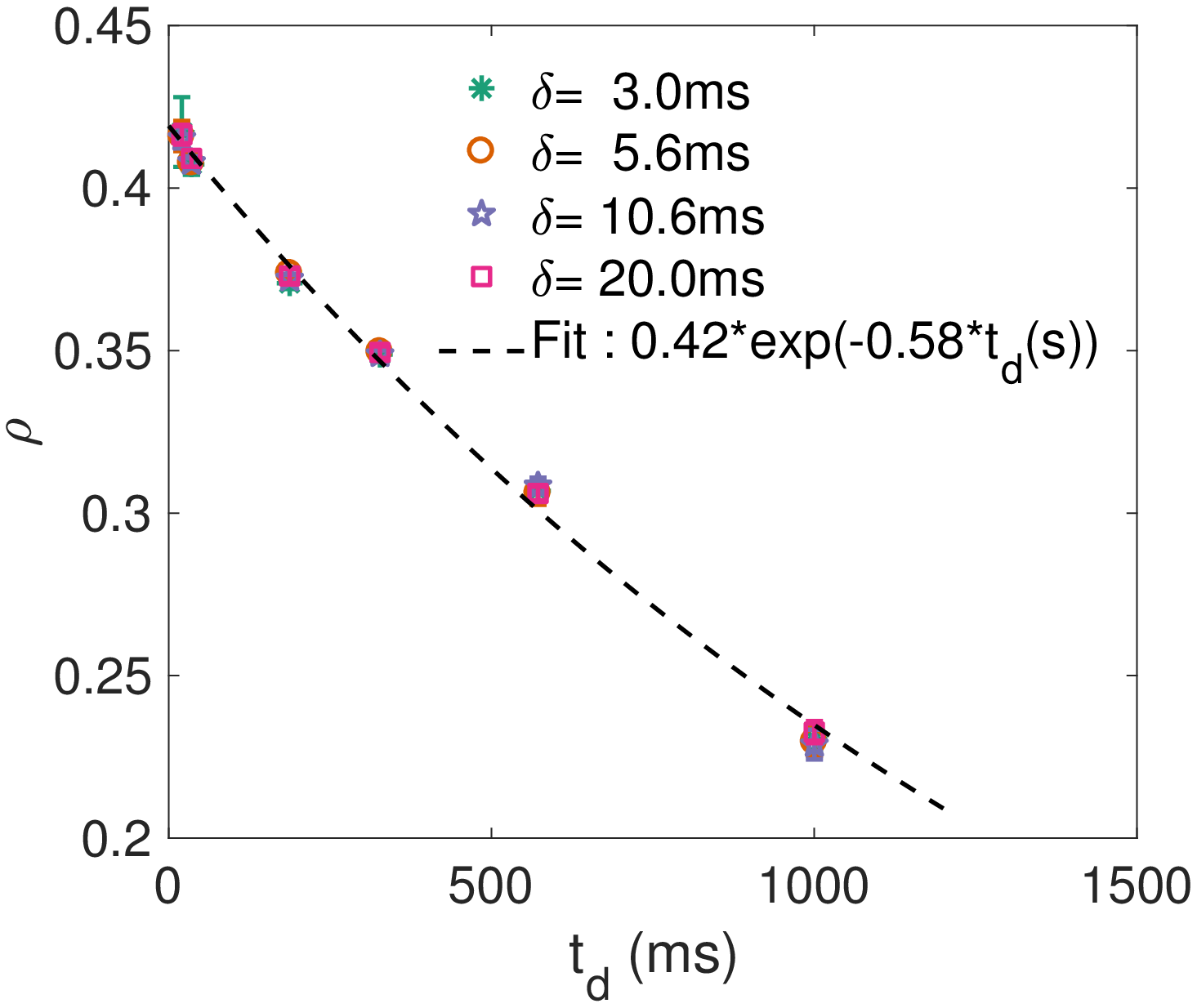}
\includegraphics[width=50mm]{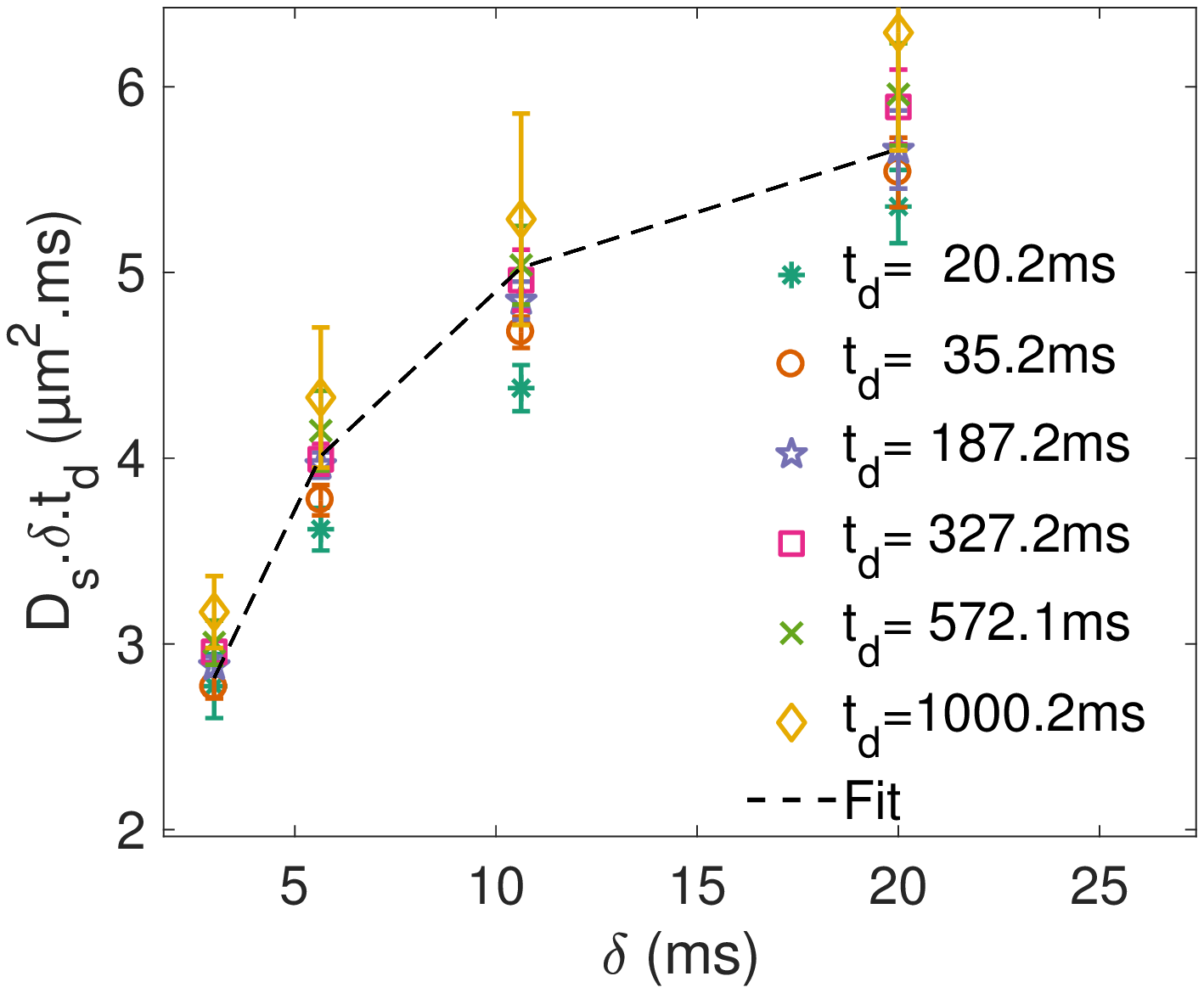}
\end{center}
\caption{Parameters obtained from the bi-exponential fit \labelcref{eq:biexp_no_exchange}. (left) The fast ADC, $D_f$, as a function of the diffusion time $t_d$; (center) The intracellular water fraction $\rho$ as a function of $t_d$. Dashed line shows an exponential fit \eqref{eq:rho_decay}, with $\rho_0=0.42$ and $\tau=1700\tunit$; (right) The product $D_s\delta t_d$ as a function of $\delta$. Dashed line shows a fit of the curves by Eq. \eqref{eq:D_s_mod}.
}
\label{fig:biexp}
\end{figure*}

The intracellular water fraction $\rho$ does not depend on $\delta$ and decreases with $t_d$, from $0.42$ at $t_d=20.2\tunit$ to $0.23$ at $t_d=1000\tunit$, and the exponential decay \eqref{eq:rho_decay} fits well (\cref{fig:biexp}), from which we estimate a typical leakage time $\tau_{i\to e}$ of about $1700 \pm 100\tunit$. Moreover the intracellular water fraction $\rho_0$ is equal to $0.42\pm 0.01$, that yields $\tau_K=\tau_{i\to e}/\rho_0 = 4000\pm 300\tunit$. Note that the hypothesis $\delta\ll\tau_{i\to e}$ is valid.

The fast diffusion coefficient $D_f$ does not depend on $\delta$ and slowly decreases with $t_d$, from $1.6\dunit$ at $t_d=20.2\tunit$ to $1.2\dunit$ at $t_d=1000\tunit$ (\cref{fig:biexp}). 
As explained previously, one can interpret this decrease as the combined effect of hindered diffusion due to the high concentration of yeast cells and exchange with intracellular water. In \cite{Swiet1996a} an asymptotic formula for the time dependent diffusion coefficient in a \emph{dilute} suspension of spheres was derived. This formula indicates that the diffusion coefficient decreases towards a limit value as $t_d^{-1}$ with a typical time scale given by $R^2/D_e$, which in our case is equal to about $5\tunit$. In a \emph{crowded} suspension one expects this time scale to be linked to some correlation length of the distribution of the cells. For example, if the cells aggregate and form clusters of size $L\gg R$, $D_f$ will decrease with a time scale $L^2/D_e \gg R^2/D_e$. 
Numerous works have also been devoted to the infinite time limit of the diffusion coefficient outside an isotropic random suspension of spheres \cite{Maxwell1873a,Hashin1962a,Weissberg1963a,Jeffrey1973a,Brakel1974a}, with a common agreement on the upper bound:
\begin{equation}
\frac{D(t=\infty)}{D_e}\leq\frac{1-\rho}{1+\rho/2}\;,
\label{eq:D_inf}
\end{equation}
where the exact value of $D(t\!\!=\!\!\infty)/D_e$ depends on the distribution of spheres. In particular, this upper bound is reached in the case of a \textquote{well-separated} array of spheres, that is a suspension with no aggregates. In our case, $\rho \approx 0.4$ so that \cref{eq:D_inf} provides the upper bound $D(t=\infty)/D_e \leq 0.5$. The free diffusion coefficient of water at room temperature is around $2.3\dunit$ \cite{Tofts2000a,Mills1973a,Wang1965a} thus the hindered diffusion coefficient should be lower than $1.2 \dunit$. However $D_f$ is above $1.2\dunit$ even at $t_d$ as high as $1000\tunit$. As a consequence, the exchange alone does not seem to explain the obtained values of $D_f$. Note that, in general, neglecting the effect of geometrical hindrance on the time variation of $D_f$ leads to an underestimation of $\tau_K$.

The product $D_s\delta t_d$ is not exactly constant but increases with $\delta$ (its value at $\delta=20\tunit$ is approximately the double of its value at $\delta=3\tunit$) and slightly increases with $t_d$ (a $20\%$ increase from $t_d=20\tunit$ to $t_d=1000\tunit$) (\cref{fig:biexp}). The correction formula \eqref{eq:D_s_mod} accounts quite well for the variation with $\delta$ but is unable to reproduce the dependence on $t_d$ because the correction term in \cref{eq:D_s_mod} does not depend on $t_d$ if $t_d\gg\delta$ (which is the case for almost all data points). We expect that the variation with $t_d$ is caused by the exchange across the cell membranes. This dependence on $t_d$ makes hard to give precise estimates of $R$ and $D_i$. We get $R=2.6\pm1 \lunit$ and $D_i=0.75\pm0.15 \dunit$ ($95\%$ confidence intervals), in agreement with the values found in the literature \cite{Grebenkov2014a,Aaslund2009b}.

\subsection{K{\"a}rger model and modified K{\"a}rger model}

\begin{figure}[tb]
\begin{center}
\includegraphics[width=75mm]{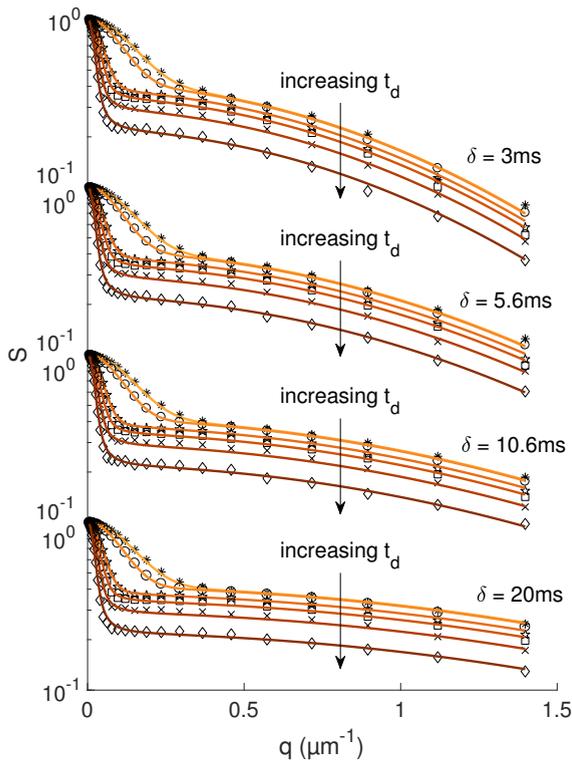}
\end{center}
\caption{Fit of the data by the modified K{\"a}rger model. The signal is plotted against $q=\gamma g \delta$ for various values of $\delta$ and $t_d$ (asterisk: $20.2\tunit$, circle: $35.2\tunit$, star: $187.2\tunit$, square: $327.2\tunit$, cross: $572.1\tunit$, diamond: $1000.2\tunit$).
Note that the plots are vertically shifted with different $\delta$ for visibility.
}
\label{fig:fit_Karger_mod}
\end{figure}

On these experimental data, the K{\"a}rger model and the modified K{\"a}rger model yield very close values of the parameters, hence we only show in Fig. \ref{fig:fit_Karger_mod} a fit made with the modified K{\"a}rger model. In spite of small systematic deviations between the data and the model, the fit is good and yields (with $95\%$ confidence intervals): $D_f=1.73\pm 0.03\dunit$, $D_i=0.86\pm 0.12\dunit$, $\rho=0.413\pm 0.002$, $\tau_K=3700 \pm 100\tunit$ and $R=2.7 \pm 0.07\lunit$. From \cref{eq:tau_kappa,eq:tau_tauK} one deduces the permeability $\kappa=(5.8\pm 0.4)\, 10^{-4}\kunit$. These values are consistent with the literature \cite{Grebenkov2014a,Aaslund2009b}. \Cref{fig:fit_Karger_mod} illustrates also the property that the low-$q$ decay of the signal is independent of $\delta$ whereas the high-$q$ decay is independent of $t_d$ (more precisely, varying $t_d$ changes only the amplitude but not the shape of the curve).

Note however that the K{\"a}rger model is only suited to fit data with several values of $t_d$ and $\delta$ at the same time. If one tries to fit only one curve $S(q)$ (that is, with one value of $t_d$ and one value of $\delta$), the fit is unstable.
Indeed, we have already noted that the bi-exponential model fits the data well (no sign of a systematic deviation, RMSE close to the noise level estimation). As a consequence, the addition of another parameter $\tau_K$ does not significantly improves the quality of the fit. Moreover, the fit algorithm returns very high values of $\tau_K$ associated with very large error bars. In turn, these large error bars on $\tau_K$ affect the stability of the whole fit because all the parameters are correlated (in particular $\rho$ and $\tau_K$).
\begin{figure}[tb]
\begin{center}
\includegraphics[width=75mm]{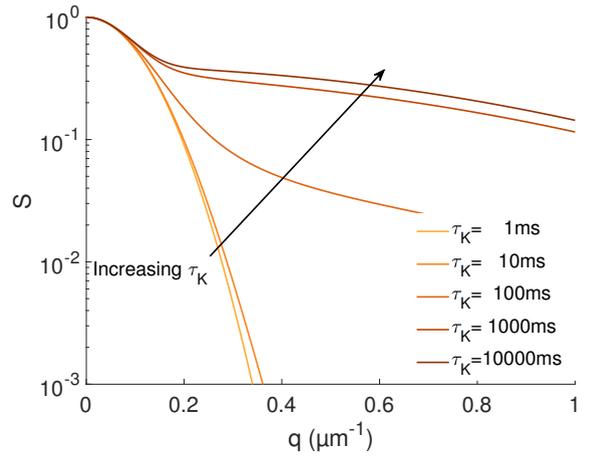}
\end{center}
\caption{The K{\"a}rger signal for $t_d=100\tunit$ and various values of $\tau_K$. While the signal increases with $\tau_K$, the dependence on $\tau_K$ is weak when $\tau_K\ll t_d$ or $\tau_K\gg t_d$.
}
\label{fig:Karger}
\end{figure}
This can be understood by looking at Fig. \ref{fig:Karger}. The signal \textcolor{black}{is sensitive to} $\tau_K$ only when $\tau_K \sim t_d$. As $\tau_K/t_d\to 0$ the signal converges to the fast mono-exponential decay and as $\tau_K/t_d \to \infty$ the signal converges to the bi-exponential decay. As the bi-exponential fit is already good, the optimal value of $\tau_K$ is high compared to $t_d$ and it is not well-determined. One also notices that the two curves with the highest $\tau_K$ ($10^3\tunit$ and $10^4\tunit$) have both the shape of a bi-exponential decay, the only difference being the apparent value of $\rho$ (the amplitude of the slow high-$q$ decay).

\section{Conclusion}

In summary, the bi-exponential model and both K{\"a}rger models yield rather close values of the parameters. In particular, the intracellular water fraction $\rho$ and the exchange time $\tau_K$ are very similar.  The bi-exponential model shows its limitations when it comes to the analysis of the slow apparent diffusion coefficient $D_s$. Indeed $D_s t_d$ weakly depends on $t_d$ whereas it should not, according to \cref{eq:D_s_mod}. This effect may be attributed to the exchange. In the same way, the slow dependence of $D_f$ on $t_d$ may be caused by the exchange as well as by the hindering by the cells. The errors bars on the parameters obtained from the bi-exponential model are also slightly larger than the ones obtained from the K{\"a}rger model.

The modified K{\"a}rger model is the most appropriate one from a theoretical point of view and can fit the whole data with one set of parameters. In some sense, this strength is also a weakness because the model is not applicable if one does not have full sets of data with variable $q$ and $t_d$. Furthermore, this makes the model too \textquote{rigid}; for example, it is not clear how to take into account time-dependent diffusion coefficients.

On the other hand, the bi-exponential model with time-decaying $\rho$ has a transparent physical interpretation and suggests the following experimental modality to quickly measure the exchange time: to choose a fixed value of $q$ with fixed $\delta$ and to probe the signal as a function of diffusion time, for example with a multiple echo (CPMG) experiment (this is analogous to the Cg-simulations of Ref. \cite{Meier2003a}). At short times, the signal from the extracellular water is not completely destroyed, but at long times one only measures the intracellular signal, which decays as $\exp(-t/\tau_{i\to e})$, as shown above. Note that the same measurement without any weighting gradient is also needed in order to estimate the $T_2$-relaxation beforehand. \textcolor{black}{This modality bears similarities with the FEXSY and FEXI sequences \cite{Aaslund2009a,Lasic2011b} (where an additional filtering sequence is used to destroy the extracellular signal)}. From a theoretical point of view, one should choose $\delta$ large enough in order to be in the restricted diffusion regime but small compared to $\tau_{i\to e}$. This is only possible if $\tau_{i\to e} \gtrsim 50\tunit$. Another condition is that the echo time $TE$ should be chosen sufficiently long so that the extracellular magnetization is completely destroyed between two echoes ($D_e q^2 TE \gg 1$) but still not too large compared to $\tau_{i\to e}$. On a conventional scanner with $g \leq 20 \gunit$ these conditions require that $\tau_{i\to e} \gtrsim 250\tunit$. With gradients higher than about $200 \gunit$ one can theoretically probe exchange time as short as $50\tunit$.
\bibliographystyle{model1-num-names}

\begin{thebibliography}{66}
\expandafter\ifx\csname natexlab\endcsname\relax\def\natexlab#1{#1}\fi
\providecommand{\url}[1]{\texttt{#1}}
\providecommand{\href}[2]{#2}
\providecommand{\path}[1]{#1}
\providecommand{\DOIprefix}{doi:}
\providecommand{\ArXivprefix}{arXiv:}
\providecommand{\URLprefix}{URL: }
\providecommand{\Pubmedprefix}{pmid:}
\providecommand{\doi}[1]{\href{http://dx.doi.org/#1}{\path{#1}}}
\providecommand{\Pubmed}[1]{\href{pmid:#1}{\path{#1}}}
\providecommand{\bibinfo}[2]{#2}
\ifx\xfnm\relax \def\xfnm[#1]{\unskip,\space#1}\fi
\bibitem[{Callaghan(1991)}]{Callaghan1991a}
\bibinfo{author}{P.~T. Callaghan}, \bibinfo{title}{Principles of Nuclear
  Magnetic Resonance Microscopy}, \bibinfo{edition}{1st} ed.,
  \bibinfo{publisher}{Clarendon Press}, \bibinfo{year}{1991}.
\bibitem[{Price(2009)}]{Price2009a}
\bibinfo{author}{W.~Price}, \bibinfo{title}{{NMR} Studies of Translational
  Motion: {Principles} and Applications}, \bibinfo{publisher}{Cambridge
  Molecular Science}, \bibinfo{year}{2009}.
\bibitem[{Grebenkov(2007)}]{Grebenkov2007a}
\bibinfo{author}{D.~S. Grebenkov},
\newblock \bibinfo{title}{{NMR} survey of reflected {Brownian} motion},
\newblock \bibinfo{journal}{Rev. Mod. Phys.} \bibinfo{volume}{79}
  (\bibinfo{year}{2007}) \bibinfo{pages}{1077--1137}.
\bibitem[{Tuch et~al.(2003)Tuch, Reese, Wiegell, and Wedeen}]{Tuch2003a}
\bibinfo{author}{D.~S. Tuch}, \bibinfo{author}{T.~G. Reese},
  \bibinfo{author}{M.~R. Wiegell}, \bibinfo{author}{V.~J. Wedeen},
\newblock \bibinfo{title}{Diffusion {MRI} of complex neural architecture},
\newblock \bibinfo{journal}{Neuron} \bibinfo{volume}{40} (\bibinfo{year}{2003})
  \bibinfo{pages}{885 -- 895}.
\bibitem[{Frahm et~al.(2004)Frahm, Dechent, Baudewig, and
  Merboldt}]{Frahm2004a}
\bibinfo{author}{J.~Frahm}, \bibinfo{author}{P.~Dechent},
  \bibinfo{author}{J.~Baudewig}, \bibinfo{author}{K.~Merboldt},
\newblock \bibinfo{title}{Advances in functional {MRI} of the human brain},
\newblock \bibinfo{journal}{Prog. Nucl. Magn. Reson. Spectrosc.}
  \bibinfo{volume}{44} (\bibinfo{year}{2004}) \bibinfo{pages}{1 -- 32}.
\bibitem[{Bihan and Johansen-Berg(2012)}]{Bihan2012a}
\bibinfo{author}{D.~L. Bihan}, \bibinfo{author}{H.~Johansen-Berg},
\newblock \bibinfo{title}{Diffusion {MRI} at 25: {Exploring} brain tissue
  structure and function},
\newblock \bibinfo{journal}{Neuroimage} \bibinfo{volume}{61}
  (\bibinfo{year}{2012}) \bibinfo{pages}{324 -- 341}.
\bibitem[{Stejskal and Tanner(1965)}]{Stejskal1965a}
\bibinfo{author}{E.~O. Stejskal}, \bibinfo{author}{J.~E. Tanner},
\newblock \bibinfo{title}{Spin diffusion measurements: {Spin} echoes in the
  presence of a time-dependent field gradient},
\newblock \bibinfo{journal}{J. Chem. Phys.} \bibinfo{volume}{42}
  (\bibinfo{year}{1965}) \bibinfo{pages}{288--292}.
\bibitem[{Kiselev(2017)}]{Kiselev2017a}
\bibinfo{author}{V.~G. Kiselev},
\newblock \bibinfo{title}{Fundamentals of diffusion {MRI} physics},
\newblock \bibinfo{journal}{NMR Biomed.} \bibinfo{volume}{30}
  (\bibinfo{year}{2017}) \bibinfo{pages}{e3602}.
\bibitem[{Niendorf et~al.(1996)Niendorf, Dijkhuizen, Norris, van
  Lookeren~Campagne, and Nicolay}]{Niendorf1996a}
\bibinfo{author}{T.~Niendorf}, \bibinfo{author}{R.~M. Dijkhuizen},
  \bibinfo{author}{D.~G. Norris}, \bibinfo{author}{M.~van Lookeren~Campagne},
  \bibinfo{author}{K.~Nicolay},
\newblock \bibinfo{title}{Biexponential diffusion attenuation in various states
  of brain tissue: {Implications} for diffusion-weighted imaging},
\newblock \bibinfo{journal}{Magn. Reson. Med.} \bibinfo{volume}{36}
  (\bibinfo{year}{1996}) \bibinfo{pages}{847--857}.
\bibitem[{Mulkern et~al.(1999)Mulkern, Gudbjartsson, Westin, Zengingonul,
  Gartner, Guttmann, Robertson, Kyriakos, Schwartz, Holtzman, Jolesz, and
  Maier}]{Mulkern1999a}
\bibinfo{author}{R.~V. Mulkern}, \bibinfo{author}{H.~Gudbjartsson},
  \bibinfo{author}{C.-F. Westin}, \bibinfo{author}{H.~P. Zengingonul},
  \bibinfo{author}{W.~Gartner}, \bibinfo{author}{C.~R.~G. Guttmann},
  \bibinfo{author}{R.~L. Robertson}, \bibinfo{author}{W.~Kyriakos},
  \bibinfo{author}{R.~Schwartz}, \bibinfo{author}{D.~Holtzman},
  \bibinfo{author}{F.~A. Jolesz}, \bibinfo{author}{S.~E. Maier},
\newblock \bibinfo{title}{Multi-component apparent diffusion coefficients in
  human brain},
\newblock \bibinfo{journal}{NMR Biomed.} \bibinfo{volume}{12}
  (\bibinfo{year}{1999}) \bibinfo{pages}{51--62}.
\bibitem[{Clark and Le~Bihan(2000)}]{Clark2000a}
\bibinfo{author}{C.~A. Clark}, \bibinfo{author}{D.~Le~Bihan},
\newblock \bibinfo{title}{Water diffusion compartmentation and anisotropy at
  high b values in the human brain},
\newblock \bibinfo{journal}{Magn. Reson. Med.} \bibinfo{volume}{44}
  (\bibinfo{year}{2000}) \bibinfo{pages}{852--859}.
\bibitem[{Chin et~al.(2002)Chin, Wehrli, Hwang, Takahashi, and
  Hackney}]{Chin2002a}
\bibinfo{author}{C.-L. Chin}, \bibinfo{author}{F.~W. Wehrli},
  \bibinfo{author}{S.~N. Hwang}, \bibinfo{author}{M.~Takahashi},
  \bibinfo{author}{D.~B. Hackney},
\newblock \bibinfo{title}{Biexponential diffusion attenuation in the rat spinal
  cord: {Computer} simulations based on anatomic images of axonal
  architecture},
\newblock \bibinfo{journal}{Magn. Reson. Med.} \bibinfo{volume}{47}
  (\bibinfo{year}{2002}) \bibinfo{pages}{455--460}.
\bibitem[{Sehy et~al.(2002)Sehy, Ackerman, and Neil}]{Sehy2002a}
\bibinfo{author}{J.~V. Sehy}, \bibinfo{author}{J.~J. Ackerman},
  \bibinfo{author}{J.~J. Neil},
\newblock \bibinfo{title}{Evidence that both fast and slow water {ADC}
  components arise from intracellular space},
\newblock \bibinfo{journal}{Magn. Reson. Med.} \bibinfo{volume}{48}
  (\bibinfo{year}{2002}) \bibinfo{pages}{765--770}.
\bibitem[{Ababneh et~al.(2005)Ababneh, Beloeil, Berde, Gambarota, Maier, and
  Mulkern}]{Ababneh2005a}
\bibinfo{author}{Z.~Ababneh}, \bibinfo{author}{H.~Beloeil},
  \bibinfo{author}{C.~B. Berde}, \bibinfo{author}{G.~Gambarota},
  \bibinfo{author}{S.~E. Maier}, \bibinfo{author}{R.~V. Mulkern},
\newblock \bibinfo{title}{Biexponential parameterization of diffusion and {T2}
  relaxation decay curves in a rat muscle edema model: {Decay} curve components
  and water compartments},
\newblock \bibinfo{journal}{Magn. Reson. Med.} \bibinfo{volume}{54}
  (\bibinfo{year}{2005}) \bibinfo{pages}{524--531}.
\bibitem[{Grebenkov(2010)}]{Grebenkov2010b}
\bibinfo{author}{D.~S. Grebenkov},
\newblock \bibinfo{title}{Use, misuse, and abuse of apparent diffusion
  coefficients},
\newblock \bibinfo{journal}{Conc. Magn. Res. A} \bibinfo{volume}{36A}
  (\bibinfo{year}{2010}) \bibinfo{pages}{24--35}.
\bibitem[{Schwarcz et~al.(2004)Schwarcz, Bogner, Meric, Correze, Berente, Pál,
  Gallyas, Doczi, Gillet, and Beloeil}]{Schwarcz2004a}
\bibinfo{author}{A.~Schwarcz}, \bibinfo{author}{P.~Bogner},
  \bibinfo{author}{P.~Meric}, \bibinfo{author}{J.-L. Correze},
  \bibinfo{author}{Z.~Berente}, \bibinfo{author}{J.~Pál},
  \bibinfo{author}{F.~Gallyas}, \bibinfo{author}{T.~Doczi},
  \bibinfo{author}{B.~Gillet}, \bibinfo{author}{J.-C. Beloeil},
\newblock \bibinfo{title}{The existence of biexponential signal decay in
  magnetic resonance diffusion-weighted imaging appears to be independent of
  compartmentalization},
\newblock \bibinfo{journal}{Magn. Reson. Med.} \bibinfo{volume}{51}
  (\bibinfo{year}{2004}) \bibinfo{pages}{278--285}.
\bibitem[{Kiselev and Il'yasov(2007)}]{Kiselev2007a}
\bibinfo{author}{V.~G. Kiselev}, \bibinfo{author}{K.~A. Il'yasov},
\newblock \bibinfo{title}{Is the “biexponential diffusion” biexponential?},
\newblock \bibinfo{journal}{Magn. Reson. Med.} \bibinfo{volume}{57}
  (\bibinfo{year}{2007}) \bibinfo{pages}{464--469}.
\bibitem[{Stanisz(2003)}]{Stanisz2003a}
\bibinfo{author}{G.~J. Stanisz},
\newblock \bibinfo{title}{Diffusion {MR} in biological systems: {Tissue}
  compartments and exchange},
\newblock \bibinfo{journal}{Isr. J. Chem.} \bibinfo{volume}{43}
  (\bibinfo{year}{2003}) \bibinfo{pages}{33--44}.
\bibitem[{Lee and Springer(2003)}]{Lee2003a}
\bibinfo{author}{J.-H. Lee}, \bibinfo{author}{C.~S. Springer},
\newblock \bibinfo{title}{Effects of equilibrium exchange on diffusion-weighted
  {NMR} signals: {The} diffusigraphic “shutter-speed”},
\newblock \bibinfo{journal}{Magn. Reson. Med.} \bibinfo{volume}{49}
  (\bibinfo{year}{2003}) \bibinfo{pages}{450--458}.
\bibitem[{Novikov and Kiselev(2010)}]{Novikov2010a}
\bibinfo{author}{D.~S. Novikov}, \bibinfo{author}{V.~G. Kiselev},
\newblock \bibinfo{title}{Effective medium theory of a diffusion-weighted
  signal},
\newblock \bibinfo{journal}{NMR Biomed.} \bibinfo{volume}{23}
  (\bibinfo{year}{2010}) \bibinfo{pages}{682--697}.
\bibitem[{K{\"a}rger(1969)}]{Kaerger1969a}
\bibinfo{author}{J.~K{\"a}rger},
\newblock \bibinfo{title}{Zur {Bestimmung} der {Diffusion} in einem
  {Zweibereichsystem} mit {Hilfe} von gepulsten {Feldgradienten}},
\newblock \bibinfo{journal}{Ann. Phys.} \bibinfo{volume}{479}
  (\bibinfo{year}{1969}) \bibinfo{pages}{1--4}.
\bibitem[{K{\"a}rger(1971)}]{Kaerger1971a}
\bibinfo{author}{J.~K{\"a}rger},
\newblock \bibinfo{title}{Der {Einflu{\ss}} der {Zweibereichdiffusion} auf die
  {Spinechod{\"a}mpfung} unter {Ber{\"u}cksichtigung} der {Relaxation} bei
  {Messungen} mit der {Methode} der gepulsten {Feldgradienten}},
\newblock \bibinfo{journal}{Ann. Phys.} \bibinfo{volume}{482}
  (\bibinfo{year}{1971}) \bibinfo{pages}{107--109}.
\bibitem[{K{\"a}rger(1985)}]{Kaerger1985a}
\bibinfo{author}{J.~K{\"a}rger},
\newblock \bibinfo{title}{{NMR} self-diffusion studies in heterogeneous
  systems},
\newblock \bibinfo{journal}{Adv. Colloid Interface Sci.} \bibinfo{volume}{23}
  (\bibinfo{year}{1985}) \bibinfo{pages}{129 -- 148}.
\bibitem[{K{\"a}rger et~al.(1988)K{\"a}rger, Pfeifer, and Heink}]{Kaerger1988a}
\bibinfo{author}{J.~K{\"a}rger}, \bibinfo{author}{H.~Pfeifer},
  \bibinfo{author}{W.~Heink},
\newblock \bibinfo{title}{Principles and application of self-diffusion
  measurements by nuclear magnetic resonance},
\newblock volume~\bibinfo{volume}{12} of \textit{\bibinfo{series}{Advances in
  Magnetic and Optical Resonance}}, \bibinfo{publisher}{Academic Press},
  \bibinfo{year}{1988}, pp. \bibinfo{pages}{1 -- 89}. \URLprefix
  \url{http://www.sciencedirect.com/science/article/pii/B978012025512250004X}.
  \DOIprefix\doi{10.1016/b978-0-12-025512-2.50004-x}.
\bibitem[{Melchior et~al.(2017)Melchior, Majer, and Kreuer}]{Melchior2017a}
\bibinfo{author}{J.-P. Melchior}, \bibinfo{author}{G.~Majer},
  \bibinfo{author}{K.-D. Kreuer},
\newblock \bibinfo{title}{Why do proton conducting polybenzimidazole phosphoric
  acid membranes perform well in high-temperature {PEM} fuel cells?},
\newblock \bibinfo{journal}{Physical Chemistry Chemical Physics}
  \bibinfo{volume}{19} (\bibinfo{year}{2017}) \bibinfo{pages}{601--612}.
\bibitem[{Lauerer et~al.(2018)Lauerer, Kurzhals, Toufar, Freude, and
  Kärger}]{Lauerer2018a}
\bibinfo{author}{A.~Lauerer}, \bibinfo{author}{R.~Kurzhals},
  \bibinfo{author}{H.~Toufar}, \bibinfo{author}{D.~Freude},
  \bibinfo{author}{J.~Kärger},
\newblock \bibinfo{title}{Tracing compartment exchange by {NMR} diffusometry:
  {Water} in lithium-exchanged low-silica {X} zeolites},
\newblock \bibinfo{journal}{Journal of Magnetic Resonance}
  \bibinfo{volume}{289} (\bibinfo{year}{2018}) \bibinfo{pages}{1 -- 11}.
\bibitem[{Fieremans et~al.(2010)Fieremans, Novikov, Jensen, and
  Helpern}]{Fieremans2010a}
\bibinfo{author}{E.~Fieremans}, \bibinfo{author}{D.~S. Novikov},
  \bibinfo{author}{J.~H. Jensen}, \bibinfo{author}{J.~A. Helpern},
\newblock \bibinfo{title}{{Monte Carlo} study of a two-compartment exchange
  model of diffusion},
\newblock \bibinfo{journal}{NMR Biomed.} \bibinfo{volume}{23}
  (\bibinfo{year}{2010}) \bibinfo{pages}{711--724}.
\bibitem[{Nilsson et~al.(2013)Nilsson, van Westen, St{\aa}hlberg, Sundgren, and
  L{\"a}tt}]{Nilsson2013a}
\bibinfo{author}{M.~Nilsson}, \bibinfo{author}{D.~van Westen},
  \bibinfo{author}{F.~St{\aa}hlberg}, \bibinfo{author}{P.~C. Sundgren},
  \bibinfo{author}{J.~L{\"a}tt},
\newblock \bibinfo{title}{The role of tissue microstructure and water exchange
  in biophysical modelling of diffusion in white matter},
\newblock \bibinfo{journal}{Magnetic Resonance Materials in Physics, Biology
  and Medicine} \bibinfo{volume}{26} (\bibinfo{year}{2013})
  \bibinfo{pages}{345--370}.
\bibitem[{Coatl{\'e}ven et~al.(2014)Coatl{\'e}ven, Haddar, and
  Li}]{Coatleven2014a}
\bibinfo{author}{J.~Coatl{\'e}ven}, \bibinfo{author}{H.~Haddar},
  \bibinfo{author}{J.-R. Li},
\newblock \bibinfo{title}{A macroscopic model including membrane exchange for
  diffusion {MRI}},
\newblock \bibinfo{journal}{SIAM J. Appl. Math.} \bibinfo{volume}{74}
  (\bibinfo{year}{2014}) \bibinfo{pages}{516--546}.
\bibitem[{Li et~al.(2014)Li, Nguyen, Nguyen, Haddar, Coatl{\'e}ven, and
  Bihan}]{Li2014a}
\bibinfo{author}{J.-R. Li}, \bibinfo{author}{H.~T. Nguyen},
  \bibinfo{author}{D.~V. Nguyen}, \bibinfo{author}{H.~Haddar},
  \bibinfo{author}{J.~Coatl{\'e}ven}, \bibinfo{author}{D.~L. Bihan},
\newblock \bibinfo{title}{Numerical study of a macroscopic finite pulse model
  of the diffusion {MRI} signal},
\newblock \bibinfo{journal}{J. Magn. Reson.} \bibinfo{volume}{248}
  (\bibinfo{year}{2014}) \bibinfo{pages}{54 -- 65}.
\bibitem[{Price et~al.(1998)Price, Barzykin, Hayamizu, and
  Tachiya}]{Price1998a}
\bibinfo{author}{W.~S. Price}, \bibinfo{author}{A.~V. Barzykin},
  \bibinfo{author}{K.~Hayamizu}, \bibinfo{author}{M.~Tachiya},
\newblock \bibinfo{title}{A model for diffusive transport through a spherical
  interface probed by pulsed-field gradient {NMR}},
\newblock \bibinfo{journal}{Biophys. J.} \bibinfo{volume}{74}
  (\bibinfo{year}{1998}) \bibinfo{pages}{2259 -- 2271}.
\bibitem[{Neuman(1974)}]{Neuman1974a}
\bibinfo{author}{C.~H. Neuman},
\newblock \bibinfo{title}{Spin echo of spins diffusing in a bounded medium},
\newblock \bibinfo{journal}{J. Chem. Phys.} \bibinfo{volume}{60}
  (\bibinfo{year}{1974}) \bibinfo{pages}{4508--4511}.
\bibitem[{Robertson(1966)}]{Robertson1966a}
\bibinfo{author}{B.~Robertson},
\newblock \bibinfo{title}{Spin-echo decay of spins diffusing in a bounded
  region},
\newblock \bibinfo{journal}{Phys. Rev.} \bibinfo{volume}{151}
  (\bibinfo{year}{1966}) \bibinfo{pages}{273--277}.
\bibitem[{de~Swiet and Sen(1994)}]{Swiet1994a}
\bibinfo{author}{T.~M. de~Swiet}, \bibinfo{author}{P.~N. Sen},
\newblock \bibinfo{title}{Decay of nuclear magnetization by bounded diffusion
  in a constant field gradient},
\newblock \bibinfo{journal}{J. Chem. Phys.} \bibinfo{volume}{100}
  (\bibinfo{year}{1994}) \bibinfo{pages}{5597--5604}.
\bibitem[{Hurlimann et~al.(1995)Hurlimann, Helmer, Deswiet, Sen, and
  Sotak}]{Hurlimann1995a}
\bibinfo{author}{M.~D. Hurlimann}, \bibinfo{author}{K.~G. Helmer},
  \bibinfo{author}{T.~M. Deswiet}, \bibinfo{author}{P.~N. Sen},
  \bibinfo{author}{C.~H. Sotak},
\newblock \bibinfo{title}{Spin echoes in a constant gradient and in the
  presence of simple restriction},
\newblock \bibinfo{journal}{Journal of Magnetic Resonance, Series A}
  \bibinfo{volume}{113} (\bibinfo{year}{1995}) \bibinfo{pages}{260 -- 264}.
\bibitem[{Grebenkov(2014)}]{Grebenkov2014b}
\bibinfo{author}{D.~S. Grebenkov},
\newblock \bibinfo{title}{Exploring diffusion across permeable barriers at high
  gradients. {II}. {Localization} regime},
\newblock \bibinfo{journal}{J. Magn. Reson.} \bibinfo{volume}{248}
  (\bibinfo{year}{2014}) \bibinfo{pages}{164--176}.
\bibitem[{Kiselev and Novikov(2018)}]{Kiselev2018a}
\bibinfo{author}{V.~G. Kiselev}, \bibinfo{author}{D.~S. Novikov},
\newblock \bibinfo{title}{Transverse {NMR} relaxation in biological tissues},
\newblock \bibinfo{journal}{NeuroImage}  (\bibinfo{year}{2018}).
\bibitem[{de~Swiet and Sen(1996)}]{Swiet1996a}
\bibinfo{author}{T.~M. de~Swiet}, \bibinfo{author}{P.~N. Sen},
\newblock \bibinfo{title}{Time dependent diffusion coefficient in a disordered
  medium},
\newblock \bibinfo{journal}{J. Chem. Phys.} \bibinfo{volume}{104}
  (\bibinfo{year}{1996}) \bibinfo{pages}{206--209}.
\bibitem[{Sen(2004)}]{Sen2004a}
\bibinfo{author}{P.~N. Sen},
\newblock \bibinfo{title}{Time-dependent diffusion coefficient as a probe of
  geometry},
\newblock \bibinfo{journal}{Conc. Magn. Res. A} \bibinfo{volume}{23A}
  (\bibinfo{year}{2004}) \bibinfo{pages}{1--21}.
\bibitem[{Novikov et~al.(2014)Novikov, Jensen, Helpern, and
  Fieremans}]{Novikov2014a}
\bibinfo{author}{D.~S. Novikov}, \bibinfo{author}{J.~H. Jensen},
  \bibinfo{author}{J.~A. Helpern}, \bibinfo{author}{E.~Fieremans},
\newblock \bibinfo{title}{Revealing mesoscopic structural universality with
  diffusion},
\newblock \bibinfo{journal}{PNAS} \bibinfo{volume}{111} (\bibinfo{year}{2014})
  \bibinfo{pages}{5088--5093}.
\bibitem[{Stoller et~al.(1991)Stoller, Happer, and Dyson}]{Stoller1991a}
\bibinfo{author}{S.~D. Stoller}, \bibinfo{author}{W.~Happer},
  \bibinfo{author}{F.~J. Dyson},
\newblock \bibinfo{title}{Transverse spin relaxation in inhomogeneous magnetic
  fields},
\newblock \bibinfo{journal}{Phys. Rev. A} \bibinfo{volume}{44}
  (\bibinfo{year}{1991}) \bibinfo{pages}{7459--7477}.
\bibitem[{Callaghan et~al.(1991)Callaghan, Coy, MacGowan, Packer, and
  Zelaya}]{Callaghan1991b}
\bibinfo{author}{P.~T. Callaghan}, \bibinfo{author}{A.~Coy},
  \bibinfo{author}{D.~MacGowan}, \bibinfo{author}{K.~J. Packer},
  \bibinfo{author}{F.~O. Zelaya},
\newblock \bibinfo{title}{Diffraction-like effects in {NMR} diffusion studies
  of fluids in porous solids},
\newblock \bibinfo{journal}{Nature} \bibinfo{volume}{351}
  (\bibinfo{year}{1991}) \bibinfo{pages}{467--469}.
\bibitem[{Sapoval(1994)}]{Sapoval1994a}
\bibinfo{author}{B.~Sapoval},
\newblock \bibinfo{title}{General formulation of laplacian transfer across
  irregular surfaces},
\newblock \bibinfo{journal}{Phys. Rev. Lett.} \bibinfo{volume}{73}
  (\bibinfo{year}{1994}) \bibinfo{pages}{3314--3316}.
\bibitem[{Sapoval et~al.(2002)Sapoval, Filoche, and Weibel}]{Sapoval2002a}
\bibinfo{author}{B.~Sapoval}, \bibinfo{author}{M.~Filoche},
  \bibinfo{author}{E.~R. Weibel},
\newblock \bibinfo{title}{Smaller is better{\textemdash}but not too small: A
  physical scale for the design of the mammalian pulmonary acinus},
\newblock \bibinfo{journal}{PNAS} \bibinfo{volume}{99} (\bibinfo{year}{2002})
  \bibinfo{pages}{10411--10416}.
\bibitem[{Grebenkov(2006{\natexlab{a}})}]{Grebenkov2006c}
\bibinfo{author}{D.~S. Grebenkov},
\newblock \bibinfo{title}{Scaling properties of the spread harmonic measures},
\newblock \bibinfo{journal}{Fractals} \bibinfo{volume}{14}
  (\bibinfo{year}{2006}{\natexlab{a}}) \bibinfo{pages}{231--243}.
\bibitem[{Grebenkov(2006{\natexlab{b}})}]{Grebenkov2006d}
\bibinfo{author}{D.~Grebenkov}, \bibinfo{title}{Partially Reflected Brownian
  Motion: A Stochastic Approach to Transport Phenomena},
  \bibinfo{publisher}{Nova Science Publishers},
  \bibinfo{year}{2006}{\natexlab{b}}, pp. \bibinfo{pages}{135--169}. \URLprefix
  \url{https://www.novapublishers.com/catalog/product_info.php?products_id=3636}.
\bibitem[{Nguyen et~al.(2015)Nguyen, Grebenkov, Nguyen, Poupon, Bihan, and
  Li}]{Nguyen2015c}
\bibinfo{author}{H.~T. Nguyen}, \bibinfo{author}{D.~Grebenkov},
  \bibinfo{author}{D.~V. Nguyen}, \bibinfo{author}{C.~Poupon},
  \bibinfo{author}{D.~L. Bihan}, \bibinfo{author}{J.-R. Li},
\newblock \bibinfo{title}{Parameter estimation using macroscopic diffusion
  {MRI} signal models},
\newblock \bibinfo{journal}{Physics in Medicine \& Biology}
  \bibinfo{volume}{60} (\bibinfo{year}{2015}) \bibinfo{pages}{3389}.
\bibitem[{Ning et~al.(2018)Ning, Nilsson, Lasi{\v c}, Westin, and
  Rathi}]{Ning2018a}
\bibinfo{author}{L.~Ning}, \bibinfo{author}{M.~Nilsson},
  \bibinfo{author}{S.~Lasi{\v c}}, \bibinfo{author}{C.-F. Westin},
  \bibinfo{author}{Y.~Rathi},
\newblock \bibinfo{title}{Cumulant expansions for measuring water exchange
  using diffusion {MRI}},
\newblock \bibinfo{journal}{J. Chem. Phys.} \bibinfo{volume}{148}
  (\bibinfo{year}{2018}) \bibinfo{pages}{074109}.
\bibitem[{Meier et~al.(2003)Meier, Dreher, and Leibfritz}]{Meier2003a}
\bibinfo{author}{C.~Meier}, \bibinfo{author}{W.~Dreher},
  \bibinfo{author}{D.~Leibfritz},
\newblock \bibinfo{title}{Diffusion in compartmental systems. {I.} {A}
  comparison of an analytical model with simulations},
\newblock \bibinfo{journal}{Magn. Reson. Med.} \bibinfo{volume}{50}
  (\bibinfo{year}{2003}) \bibinfo{pages}{500--509}.
\bibitem[{Tanner(1970)}]{Tanner1970a}
\bibinfo{author}{J.~E. Tanner},
\newblock \bibinfo{title}{Use of the stimulated echo in {NMR} diffusion
  studies},
\newblock \bibinfo{journal}{J. Chem. Phys.} \bibinfo{volume}{52}
  (\bibinfo{year}{1970}) \bibinfo{pages}{2523--2526}.
\bibitem[{{\AA}slund et~al.(2008){\AA}slund, Cabaleiro-Lago, S{\"o}derman, and
  Topgaard}]{Aaslund2008a}
\bibinfo{author}{I.~{\AA}slund}, \bibinfo{author}{C.~Cabaleiro-Lago},
  \bibinfo{author}{O.~S{\"o}derman}, \bibinfo{author}{D.~Topgaard},
\newblock \bibinfo{title}{Diffusion {NMR} for determining the homogeneous
  length-scale in lamellar phases},
\newblock \bibinfo{journal}{J. Phys. Chem. B} \bibinfo{volume}{112}
  (\bibinfo{year}{2008}) \bibinfo{pages}{2782--2794}. \bibinfo{note}{PMID:
  18271569}.
\bibitem[{{\AA}slund et~al.(2011){\AA}slund, Medronho, Topgaard, S{\"o}derman,
  and Schmidt}]{Aaslund2011a}
\bibinfo{author}{I.~{\AA}slund}, \bibinfo{author}{B.~Medronho},
  \bibinfo{author}{D.~Topgaard}, \bibinfo{author}{O.~S{\"o}derman},
  \bibinfo{author}{C.~Schmidt},
\newblock \bibinfo{title}{Homogeneous length scale of shear-induced
  multilamellar vesicles studied by diffusion {NMR}},
\newblock \bibinfo{journal}{J. Magn. Reson.} \bibinfo{volume}{209}
  (\bibinfo{year}{2011}) \bibinfo{pages}{291 -- 299}.
\bibitem[{Lasi{\v{c}} et~al.(2011)Lasi{\v{c}}, {\AA}slund, Oppel, Topgaard,
  S{\"o}derman, and Gradzielski}]{Lasic2011a}
\bibinfo{author}{S.~Lasi{\v{c}}}, \bibinfo{author}{I.~{\AA}slund},
  \bibinfo{author}{C.~Oppel}, \bibinfo{author}{D.~Topgaard},
  \bibinfo{author}{O.~S{\"o}derman}, \bibinfo{author}{M.~Gradzielski},
\newblock \bibinfo{title}{Investigations of vesicle gels by pulsed and
  modulated gradient {NMR} diffusion techniques},
\newblock \bibinfo{journal}{Soft Matter} \bibinfo{volume}{7}
  (\bibinfo{year}{2011}) \bibinfo{pages}{3947--3955}.
\bibitem[{Eriksson et~al.(2017)Eriksson, Elbing, S{\"o}derman,
  Lindkvist-Petersson, Topgaard, and Lasi{\v{c}}}]{Eriksson2017a}
\bibinfo{author}{S.~Eriksson}, \bibinfo{author}{K.~Elbing},
  \bibinfo{author}{O.~S{\"o}derman}, \bibinfo{author}{K.~Lindkvist-Petersson},
  \bibinfo{author}{D.~Topgaard}, \bibinfo{author}{S.~Lasi{\v{c}}},
\newblock \bibinfo{title}{{NMR} quantification of diffusional exchange in cell
  suspensions with relaxation rate differences between intra and extracellular
  compartments},
\newblock \bibinfo{journal}{PLoS One} \bibinfo{volume}{12}
  (\bibinfo{year}{2017}) \bibinfo{pages}{1--18}.
\bibitem[{{\AA}slund and Topgaard(2009)}]{Aaslund2009b}
\bibinfo{author}{I.~{\AA}slund}, \bibinfo{author}{D.~Topgaard},
\newblock \bibinfo{title}{Determination of the self-diffusion coefficient of
  intracellular water using {PGSE} {NMR} with variable gradient pulse length},
\newblock \bibinfo{journal}{J. Magn. Reson.} \bibinfo{volume}{201}
  (\bibinfo{year}{2009}) \bibinfo{pages}{250 -- 254}.
\bibitem[{Maxwell(1873)}]{Maxwell1873a}
\bibinfo{author}{J.~C. Maxwell}, \bibinfo{title}{A treaty on eletricity and
  magnetism}, volume~\bibinfo{volume}{1}, \bibinfo{edition}{2nd} ed.,
  \bibinfo{publisher}{Clarendon Press}, \bibinfo{year}{1873}.
\bibitem[{Hashin and Shtrikman(1962)}]{Hashin1962a}
\bibinfo{author}{Z.~Hashin}, \bibinfo{author}{S.~Shtrikman},
\newblock \bibinfo{title}{A variational approach to the theory of the effective
  magnetic permeability of multiphase materials},
\newblock \bibinfo{journal}{J. Appl. Phys.} \bibinfo{volume}{33}
  (\bibinfo{year}{1962}) \bibinfo{pages}{3125--3131}.
\bibitem[{Weissberg(1963)}]{Weissberg1963a}
\bibinfo{author}{H.~L. Weissberg},
\newblock \bibinfo{title}{Effective diffusion coefficient in porous media},
\newblock \bibinfo{journal}{J. Appl. Phys.} \bibinfo{volume}{34}
  (\bibinfo{year}{1963}) \bibinfo{pages}{2636--2639}.
\bibitem[{Jeffrey(1973)}]{Jeffrey1973a}
\bibinfo{author}{D.~J. Jeffrey},
\newblock \bibinfo{title}{Conduction through a random suspension of spheres},
\newblock \bibinfo{journal}{Proc. Roy. Soc. Lond. A} \bibinfo{volume}{335}
  (\bibinfo{year}{1973}) \bibinfo{pages}{355--367}.
\bibitem[{van Brakel and Heertjes(1974)}]{Brakel1974a}
\bibinfo{author}{J.~van Brakel}, \bibinfo{author}{P.~Heertjes},
\newblock \bibinfo{title}{Analysis of diffusion in macroporous media in terms
  of a porosity, a tortuosity and a constrictivity factor},
\newblock \bibinfo{journal}{Int. J. Heat Mass Transfer} \bibinfo{volume}{17}
  (\bibinfo{year}{1974}) \bibinfo{pages}{1093--1103}.
\bibitem[{Tofts et~al.(2000)Tofts, Lloyd, Clark, Barker, Parker, McConville,
  Baldock, and Pope}]{Tofts2000a}
\bibinfo{author}{P.~Tofts}, \bibinfo{author}{D.~Lloyd},
  \bibinfo{author}{C.~Clark}, \bibinfo{author}{G.~Barker},
  \bibinfo{author}{G.~Parker}, \bibinfo{author}{P.~McConville},
  \bibinfo{author}{C.~Baldock}, \bibinfo{author}{J.~Pope},
\newblock \bibinfo{title}{Test liquids for quantitative {MRI} measurements of
  self-diffusion coefficient in vivo},
\newblock \bibinfo{journal}{Magn. Reson. Med.} \bibinfo{volume}{43}
  (\bibinfo{year}{2000}) \bibinfo{pages}{368--374}.
\bibitem[{Mills(1973)}]{Mills1973a}
\bibinfo{author}{R.~Mills},
\newblock \bibinfo{title}{Self-diffusion in normal and heavy water in the range
  1-45°},
\newblock \bibinfo{journal}{J. Phys. Chem.} \bibinfo{volume}{77}
  (\bibinfo{year}{1973}) \bibinfo{pages}{685--688}.
\bibitem[{Wang(1965)}]{Wang1965a}
\bibinfo{author}{J.~H. Wang},
\newblock \bibinfo{title}{Self-diffusion coefficients of water},
\newblock \bibinfo{journal}{J. Phys. Chem.} \bibinfo{volume}{69}
  (\bibinfo{year}{1965}) \bibinfo{pages}{4412--4412}.
\bibitem[{Grebenkov et~al.(2014)Grebenkov, Nguyen, and Li}]{Grebenkov2014a}
\bibinfo{author}{D.~S. Grebenkov}, \bibinfo{author}{D.~V. Nguyen},
  \bibinfo{author}{J.-R. Li},
\newblock \bibinfo{title}{Exploring diffusion across permeable barriers at high
  gradients. {I}. {Narrow} pulse approximation},
\newblock \bibinfo{journal}{J. Magn. Reson.} \bibinfo{volume}{248}
  (\bibinfo{year}{2014}) \bibinfo{pages}{153--163}.
\bibitem[{{\AA}slund et~al.(2009){\AA}slund, Nowacka, Nilsson, and
  Topgaard}]{Aaslund2009a}
\bibinfo{author}{I.~{\AA}slund}, \bibinfo{author}{A.~Nowacka},
  \bibinfo{author}{M.~Nilsson}, \bibinfo{author}{D.~Topgaard},
\newblock \bibinfo{title}{Filter-exchange {PGSE} {NMR} determination of cell
  membrane permeability},
\newblock \bibinfo{journal}{J. Magn. Reson.} \bibinfo{volume}{200}
  (\bibinfo{year}{2009}) \bibinfo{pages}{291 -- 295}.
\bibitem[{Lasi{\v{c}} et~al.(2011)Lasi{\v{c}}, Nilsson, L{\"a}tt,
  St{\aa}hlberg, and Topgaard}]{Lasic2011b}
\bibinfo{author}{S.~Lasi{\v{c}}}, \bibinfo{author}{M.~Nilsson},
  \bibinfo{author}{J.~L{\"a}tt}, \bibinfo{author}{F.~St{\aa}hlberg},
  \bibinfo{author}{D.~Topgaard},
\newblock \bibinfo{title}{Apparent exchange rate mapping with diffusion {MRI}},
\newblock \bibinfo{journal}{Magn. Reson. Med.} \bibinfo{volume}{66}
  (\bibinfo{year}{2011}) \bibinfo{pages}{356--365}.

\end{thebibliography}

\end{document}